# Resonant Dyakonov–Shur Magnetoplasmons in Graphene Terahertz Photodetectors

*Juan A. Delgado-Notario[1,2][⊥]\*, Cedric Bray[3][⊥], Elsa Perez-Martin[3], Ben Benhamou-Bui[3], Namrata Saha[1,2], Sahil Parvez[1,2], Christophe.Consejo[3], Guillaume Sigu[3], Salah Benlemqwanssa[3], Laurent Bonnet[3], Takashi Taniguchi[4], Kenji Watanabe[5], José M. Caridad[1,2], Sergey S. Krishtopenko[3], Yahya M. Meziani[1], Benoit Jouault[3], Jérémie Torres[3], Sandra Ruffenach[3] and Frédéric Teppe[3]\**

[1] Departamento de Física Aplicada, Universidad de Salamanca, 37008 Salamanca, Spain

[2] Unidad de Excelencia en Luz y Materia Estructurada (LUMES), Universidad de Salamanca, Salamanca 37008, Spain

[3] L2C (UMR 5221), Université de Montpellier, CNRS, Montpellier, France

[4] Research Center for Materials Nanoarchitectonics, National Institute for Materials Science, 1-1 Namiki, Tsukuba 305-0044, Japan

[5] Research Center for Electronic and Optical Materials, National Institute for Materials Science, 1-1 Namiki, Tsukuba 305-0044, Japan

\* juanandn@usal.es, frederic.teppe@umontpellier.fr

[⊥] These authors contributed equally

**Abstract**

Graphene plasmons confine incident terahertz fields far below the diffraction limit and, when hosted by a gate-defined Fabry–Perot cavity, enable electrically tunable, frequency-selective photodetectors. In a magnetic field, these plasmons hybridize with the cyclotron motion to form magnetoplasmons, offering a platform for fundamental studies and for nonreciprocal, spectrally selective, ultrasensitive terahertz photonics. However, implementing magnetoplasmon-assisted resonant transistors at terahertz frequencies has remained challenging so far. Here we use gate-dependent, on-chip terahertz photocurrent spectroscopy combined with a perpendicular magnetic field to resolve and probe the evolution of resonant magnetoplasmons in antenna-coupled monolayer and bilayer graphene TeraFETs. In monolayer graphene the dispersion reflects the Dirac nature of the carriers, exhibiting a non-monotonic density dependence due to the interplay of plasma resonance and cyclotron motion, with an inflection point at maximal plasmon–cyclotron coupling. In contrast, in bilayer graphene we recover and map a magnetoplasmon dispersion consistent with the conventional Schrödinger-type picture. These results establish graphene TeraFET devices as a robust on-chip platform for resonant magnetoplasmonics at terahertz frequencies, enabling magnetically programmable, frequency-selective photonics and opening avenues toward photodetectors with enhanced sensitivity.

**Introduction**

The ability to control light–matter interactions at the nanoscale in the terahertz (THz) range has attracted significant attention at the interface of fundamental research and device technology[1]. A particularly powerful pathway is offered by plasmons — coherent charge-density oscillations on a conductor/dielectric interface that compress incident fields deep below the diffraction limit and propagate along the surface — promising transformative advances in the THz-band for applications in the fields of spectroscopy, emission, sensing, bioimaging, chemical detection, and ultra-broadband wireless communications[2–5]. A groundbreaking route to harness these light–matter interactions in electronic platforms was theoretically proposed three decades ago by M. Dyakonov and M. Shur[6,7], suggesting that field-effect transistors operating at THz frequencies (TeraFETs) can exhibit resonant response to the incident electromagnetic radiation at the eigenfrequencies of plasmonic oscillations of the two-dimensional electron gas (2DEG) in the device's channel. In these devices, the gated channel creates a tunable plasmonic cavity whose discrete modes are set by device length, carrier density, and boundary conditions. This challenging regime, so-called resonant operation scenario, is characterized by a quality factor $Q = \omega\tau \gg 1$ (where $\omega = 2\pi f$ is the angular frequency of the incident radiation and $\tau$ the momentum relaxation time of the charge carriers in the system). Such conditions are rarely met in most electronic systems, in sharp contrast to the more common overdamped regime $Q \ll 1$, where plasmons are heavily damped in the channel and decay before even reaching the other side of the plasmonic cavity. Since this theoretical proposal, many materials have been explored to exploit plasmons in TeraFETs, most often in the overdamped (non-resonant) regime yielding broadband rectification. Among others, these studies included research on conventional semiconductor 2DEG systems such as silicon[8], GaAs[9], GaN[10], HgCdTe[11] till more recently the family of 2D materials[12–16]. By contrast, unequivocal resonant operation long remained a "pipe dream" till very recent and unique demonstrations using high-quality exfoliated bilayer and single-layer graphene TeraFETs[17,18], due to the extraordinary mobility in these systems acting as a model platform for the Dyakonov-Shur formalism and paving the way towards spectrally selective on-chip THz detectors, mixers and modulators.

Despite substantial progress in recent years, research on plasma-wave-assisted resonant operation in graphene-based TeraFETs has, to date, largely focused on tuning plasmons via excitation frequency, channel length, temperature, and carrier density[17,18], while the magnetic degree of freedom has remained essentially unexplored. Interestingly, when subjected to a perpendicular magnetic field, plasmons, irrespective of the host material or device architecture, can couple with cyclotron resonance to form hybrid modes known as magnetoplasmons. These modes result from the interplay between collective charge oscillations and the cyclotron motion of charge carriers and offer unique opportunities for investigating material properties as their dispersion encodes information on carrier effective mass, electronic interactions, and hydrodynamic effects in electron liquids[19,20], enabling appealing THz photonic applications, such as nonreciprocal plasmonic platforms[21] and ultrasensitive label-free biosensors[22]. So far, the existence of graphene magnetoplasmons and their dynamics have been studied across a totally different range of device geometries and/or experimental approaches. For instance, THz Time-domain spectroscopy (THz-TDS) and FTIR transmission measurements have revealed the presence of magnetoplasmons in epitaxial graphene structures and nanoribbons[23,24], highlighting the impact of impurities and confinement effects on plasmon frequencies[20]. Furthermore, studies have highlighted the existence of magnetoplasmonic modes at the edges of graphene structures[25,26], confirming the presence of distinct edge and bulk modes in these systems. Also, by incorporating periodic nanostructures of disks, nanoribbons or split rings, it becomes possible to engineer active metasurfaces capable of dynamically modulating the THz transmission, absorption and/or polarization spectral features, while also enabling the study of the optical nonlinearities of graphene magnetoplasmons[27,28]. Nevertheless, these studies fall outside the Dyakonov–Shur resonant plasma-wave framework, leaving the role of magnetic field on cavity-confined graphene plasmons essentially unexplored.

In this work, we combine THz photocurrent spectroscopy (discrete frequencies between 0.25 and 3.8 THz) with gate-controlled carrier density to probe the evolution of the plasma-wave assisted resonant regime in graphene TeraFETs under the presence of magnetic fields and demonstrate the coupling between Dyakonov-Shur standing plasma waves in the gated channel region to the cyclotron resonance. To do that, we used high-quality mono- and bilayer graphene encapsulated between hexagonal boron nitride (hBN) flakes to provide the clean environment necessary to excite and support long-life graphene plasmons as demonstrated in recent works[17,18]. Using single top-gated antenna-coupled TeraFETs, THz photocurrent spectroscopy technique enables the systematic study of magnetoplasma resonance modes as a function of frequency, carrier density, and magnetic field, for both massless Dirac fermions in monolayer graphene and massive Dirac fermions with parabolic dispersion in bilayer graphene. This study provides deeper insight into the behaviour of magnetoplasmons in graphene, particularly in the Dyakonov-Shur regime of standing plasma wave confined in the plasmonic cavity under the gate. In addition, our results pave the way for future investigations involving magnetoplasma waves in more complex device architectures, with potential applications in magnetoplasmonic interferometry[29,30], exploring the hydrodynamic properties of 2D Dirac plasmons[31] under magnetic field and magnetically tunable, frequency-selective photodetectors with enhanced responsivity[32].

**Methods**

To qualitatively probe the behavior of resonant graphene magnetoplasmons subjected to an incident THz radiation in the presence of magnetic field (See **Figure 1a**), our graphene TeraFET devices studied in this work were fabricated from mechanical exfoliated single-layer (Dirac-type) and bilayer (Schrödinger-type) graphene encapsulated in between relatively thin hBN flakes using a van der Waals dry-assembling technique. All the graphene based-TeraFETs in this study were defined to have a short-channel ($L_{ch}$ = 6 µm) with side-contacts for the drain and source electrodes. A metal top-gate electrode covering most of the TeraFET channel ($L_{TG}$ = 4.8 $\mu$m) was subsequently patterned to create the plasmonic cavity. To couple the incident electromagnetic waves and enable the excitation of resonant THz standing plasma waves below the gated section of the channel, a coupled bow-tie antenna was integrated between the top gate and the source electrode (See **Figure 1b**). More information about the fabrication process can be found in ***Supporting Information Note 1***.

The devices were placed inside a cryogenic magnet system (up to 6 T) with optical access and cooled down to a temperature of 1.7 K. All results presented in this work were obtained at this operation temperature unless otherwise indicated. THz Photoresponse measurements (**Figure 1a**) were performed at zero source-drain bias at different carrier densities, under a wide range of THz excitation wavelengths and/or in the presence of magnetic fields ranging up to 1.5 T (Here the magnetic field was oriented perpendicular to the surface of the sample and parallel to the wave-vector of the incident radiation). Carrier density was tuned in our TeraFETs using a voltage generator to control the applied voltage on the top-gate electrode. In order to study the photoresponse of our detectors at different frequencies, we used two radiation sources. Low THz measurements were carried out using a Schottky diode-based frequency multiplier source generating tunable THz wavelenghts from 0.220 to 0.330 THz with average output power around 10 mW. For higher THz frequencies, we used a quantum cascade laser to illuminate the sample at fixed frequencies of 2.5, 2.9, 3.3, 3.5 and 3.8 THz (optical power are about 2.5, 4, 5, 16 and 12 mW respectively). The setup includes gold parabolic mirrors to focus the incident THz beam on the sample, with the THz radiation linearly polarized parallel to the bow-tie antenna axis. Finally, the photocurrent signal was detected as a voltage drop by using a low-noise current-to-voltage preamplifier and measured via a standard lock-in technique (THz waves were electrically modulated at a frequency of 337 Hz).

**Results and discussion**

Initially, we measured the magenetotransport characteristics of our single and bilayer graphene TeraFETs via electrical measurements. Average mobilities, $\mu$, in the devices exceeded 70000 cm$^2$V$^{-1}$s$^{-1}$ for the single-layer and bilayer graphene TeraFETs at 1.7K. The momentum-relaxing scattering time of charge carriers $\tau$, can be calculated with the relation $\tau = m\mu/e$ (where $e$ is the elementary charge and $m$ is the effective mass of carriers in graphene) with the values lying above 0.4 ps for both graphene systems.

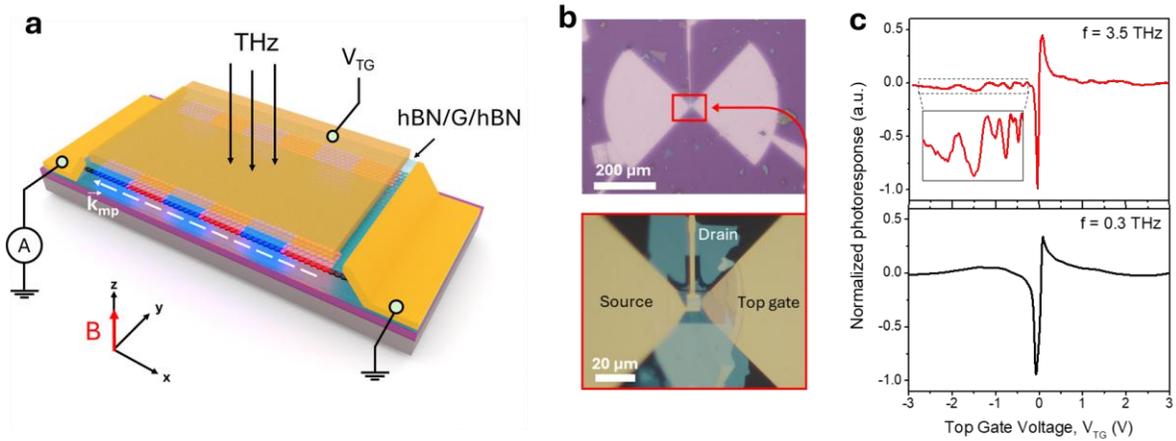

**Figure 1. Graphene-based TeraFET photodetector. a**, Schematic 3D view of the THz photocurrent measurements of the device in the presence of magnetic field highlighting the presence of resonant graphene magnetoplasmons under the gate (Plasmon wave is sketched in blue and red colors). **b**, Optical microscope image of the single-layer graphene THz detector, featuring a bow-tie antenna integrated between the source and top-gate electrodes. The bottom panel provides a zoomed-in view of the device, with Source, Top gate, and Drain contacts labeled. **c**, Nonresonant (bottom) and resonant (top) normalized THz photoresponse as a function of the top gate voltage, $V_{TG}$, measured for the monolayer TeraFET at 0.3 THz and 3.5 THz respectively for a temperature of 1.7 K and zero magnetic field.

We then studied the THz photoresponse in our devices at zero magnetic field. Firstly, the devices were illuminated at frequencies between 0.246 and 0.309 THz, intentionally operating in the non-resonant (broadband) scenario ($Q \ll 1$). As expected for the non-resonant operation regime, the photoresponse in single layer TeraFET (**Figure 1c – bottom panel**) and bilayer TeraFET (***Supporting Information Note 2***) exhibit the characteristic antisymmetric shape, reversing sign at the charge neutrality point (CNP). Apart from this central feature, the signal remains monotonic, vanishing at large gate voltages, consistent with a non-resonant overdamped regime. Near the CNP, we have observed a rectified voltage response that is slightly larger on the hole side compared to the electron side, reflecting the superior hole carrier mobility in our device.

The scenario drastically changes when the excitation frequency is increased into the high THz range, as the resonant condition ($Q > 1$) can be clearly satisfied in our graphene TeraFETs. For example, clear signatures corresponding to the appealing resonant operation in the single-layer graphene TeraFET at 3.5 THz appear on the THz photoresponse curve with prominent oscillations arising at relatively large gate voltages and being dependent on the carrier density in the system and the excitation frequency (see **Figure 1c top panel**). These peaks stem from plasmon resonances in the graphene channel, where the channel under the gate acts as a Fabry-Perot resonant cavity causing reflections at the boundaries and giving rise to standing plasma waves through constructive interferences of counter-propagating modes. Their emergences mark the transition to a resonant regime, in contrast to the non-resonant behaviour observed at lower frequencies. For completeness, we additionally measured the THz photoresponse in both systems, single-layer (**Figure 2a**) and bilayer (**Figure 2b**) TeraFETs, within the range of 2.5 THz – 3.8 THz, all within the resonant regime or weakly damped scenario ($Q > 1$). For simplicity and clarity, we only present the photoresponse evolution close to the CNP at the electron side. In this framework, excited plasmons in two-dimensional systems are characterized by their angular frequency, $\omega_p$, which can be written as:

$$\omega_p = sk \tag{1}$$

where $k$ is the real part of the wave vector[17,18] and $s$ is the plasma wave velocity defined as

$$s = \sqrt{\frac{e}{m_*}|V_{TG}|} \tag{2}$$

with $m^*$ being the effective electron mass, $e$ the elementary carrier charge and $V_{TG}$ the top gate voltage. The multiple peaks correspond to the different resonant modes occurring when the wavenumber satisfies the following condition:

$$k = \frac{\pi}{2L_{TG}}(2N + 1), N = 0, 1, 2, \ldots \tag{3}$$

here, $L_{TG}$ is the length of the gated channel (plasmonic Fabry-Perot cavity), and $N$ denotes the resonance mode number. Importantly, the effective mass $m^*$ in Eq. (2) must be carefully defined, as its value is different in single layer or bilayer graphene. On one side, $m^*$ is defined in a single-layer graphene[15,18,33] as

$$m^* = \frac{\hbar}{v_F}\sqrt{\frac{\pi C_{TG}|V_{TG}|}{e}} \tag{4}$$

with $C_{TG}$ being the thin-hBN top-gate capacitance per unit area (calculated value for our TeraFET system is 0.00112 F.m$^{-2}$, see **Supporting Information Note 3**), and $v_F$ the Fermi velocity of the charge carriers ($1.19 \cdot 10^6$ m·s$^{-1}$, as described by Chae et al[34]). In contrast, for the bilayer graphene, following the conventional Schrödinger-based picture, experimental data were fitted with an effective mass of ~0.026 $m_0$, which agrees well with values reported in the literature[35]. As a result, the plasma frequency ($f_p = \frac{\omega_p}{2\pi}$) in monolayer graphene-based TeraFETs takes the form:

$$f_p^{mono} = \frac{1}{4L_{TG}}(2N + 1)\sqrt{\frac{v_F}{\hbar}}\left(\frac{e^3|V_{TG}|}{\pi C_{TG}}\right)^{1/4} \tag{5}$$

while for the bilayer graphene TeraFETs, the plasma frequency takes the form:

$$f_p^{bi} = \frac{1}{4L_{TG}}(2N + 1)\sqrt{\frac{e}{m^*}|V_{TG}|} \tag{6}$$

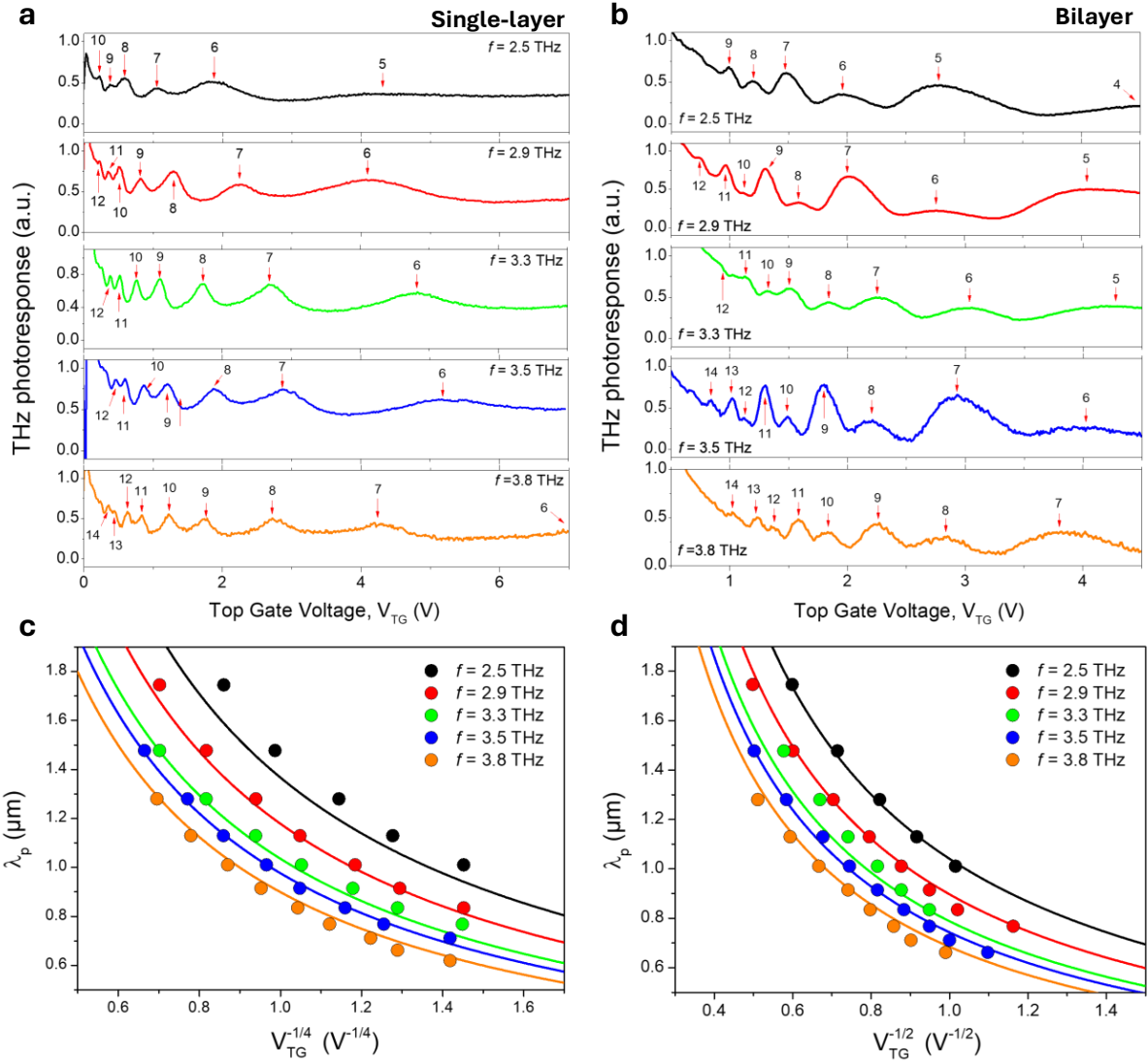

**Figure 2: Onset of plasma wave resonant regime.** Photoresponse versus top gate voltage measured at frequencies ranging from 2.5 THz to 3.8 THz, performed at 1.7K for monolayer, **a**, and bilayer, **b**, graphene based TeraFETs respectively. Resonant plasmon mode peaks, marked by red arrows, are labeled by using Eqs. (3), (5)-(6). Plasmon wavelengths, $\lambda_p$, are plotted as a function of $|V_{TG}|^{-1/4}$ for monolayer device, **c**, and a function of $|V_{TG}|^{-1/2}$ for bilayer device, **d**, for illumination frequencies at 2.5, 2.9, 3.3, 3.5 and 3.8 THz with respective symbols. $\lambda_p = 2\pi/k$ has been calculated based on the top gate voltages values where the peaks emerged in panels **b** and **c** by using Eqs. (5)-(6). The observed trends show good agreement with the Dyakonov–Shur theory adapted to monolayer graphene and bilayer graphene (solid lines).

The resonant peaks observed in **Figures 2a and 2b** correspond to modes defined by Eqs. (5)-(6) for the monolayer and bilayer graphene respectively, with their order set by the plasmonic cavity length (top gate length) and the applied gate voltage $V_{TG}$. As the frequency increases, more plasmon modes become accessible, since higher frequencies correspond to shorter wavelengths that can fit within the same channel length[17,18]. From the modal structure, the plasmon wavelength can be extracted using $\lambda_p = 2\pi/k$[17,18]. **Figures 2c and 2d** show the extracted wavelength as a function of $|V_{TG}|^{-1/4}$ for the monolayer graphene and $|V_{TG}|^{-1/2}$ for the bilayer graphene respectively, in good agreement with the expected behaviour for both systems. Similarly, one can plot the N mode number of the observed plasmons as a function of $|V_{TG}|^{-1/4}$ for the monolayer graphene and $|V_{TG}|^{-1/2}$ for the bilayer graphene highlighting a clear linear dispersion in both systems (see *Supporting Information Note 4*). The observed evolution exhibits unambiguous signatures of resonant operation consistent with Dyakonov–Shur plasma-wave theory[6,17,18,33], demonstrating the presence of resonant plasma modes in the gated plasmonic cavity and enabling a direct comparison between monolayer and bilayer graphene systems—materials with distinct plasmon dispersions—under identical excitation and boundary conditions. Beyond dispersion, the resonance linewidth also provides access to the plasmon lifetime, $\tau_p$, which can be extracted from Lorentzian fits to the observed peaks in the THz photoresponse curves. The obtained values (typically above 1 ps) slightly exceed those found in the transport times, indicating that momentum scattering alone does not set the linewidth (see *Supporting Information Note 5*).

Building on the zero-field measurements presented above, we have investigated the effect of a perpendicular variable magnetic field on the plasmonic resonances. For a 3.5 THz illumination, increasing the magnetic field causes a clear shift of the resonance peaks toward lower carrier densities (i.e. lower gate voltage), as shown in **Figure 3** and **Figure S6** (see *Supporting Information Note 6*). This shift is more pronounced for higher-order modes than for those near the CNP, indicating a mode-dependent sensitivity to the magnetic field. Such behaviour results from the coupling between plasma waves and the cyclotron motion of charge carriers, leading to the formation of magnetoplasmon modes. At higher magnetic fields, additional oscillations emerge near the CNP, consistent with Shubnikov–de Haas (SdH) oscillations, which become increasingly prominent and progressively obscure the magnetoplasmon peaks.

In the long-wavelength limit, the dispersion of two-dimensional magnetoplasmons can be expressed with the following relation[36]:

$$\omega_{mp} = \sqrt{\omega_p^2 + \omega_c^2} \qquad (7)$$

where $\omega_c$ is the cyclotron angular frequency given by $eB/(m*)$, and $\omega_p$ is the plasma-wave frequency at zero magnetic field, determined by Eqs (5) or (6) for systems with linear or parabolic bands respectively. At such, according to the Dyakonov–Shur theory presented above and adapted to monolayer graphene, the frequency of the Dirac magnetoplasmon resonance takes the following form:

$$f_{mp}^D = \frac{1}{2\pi}\sqrt{\frac{e^{3/2}v_F}{\hbar}\left(\frac{|V_{TG}|}{\pi C_{TG}}\right)^{1/2}\left(\frac{\pi}{2L_{TG}}(2N+1)\right)^2 + \left(\frac{Bv_f}{\hbar}\right)^2 \frac{e^3}{\pi C_{TG}|V_{TG}|}} \qquad (8)$$

The gate-voltage dependence of the effective mass given in Eq. (4) for monolayer graphene leads to a bulk magnetoplasmon resonance frequency with two terms explicitly proportional to $V_{TG}$. In a clear contrast, the magnetoplasmon resonance frequency for the case of bilayer graphene is given by a simpler expression, derived from the Schrödinger equation with an effective mass description:

$$f_{mp}^S = \frac{1}{2\pi}\sqrt{\left(\frac{e}{m_*}|V_{TG}|\right)\left(\frac{\pi}{2L_{TG}}(2N+1)\right)^2 + \left(\frac{eB}{m_*}\right)^2} \qquad (9)$$

where the bulk magnetoplasmon resonance frequency contains only a single term dependent on $V_{TG}$. In our measurements, graphene TeraFETs are exposed to THz radiation at a fixed frequency ($f_{source}$). Thus, gate-voltage is "tuned" in order to satisfy to the bulk magnetoplasmon resonance condition $f_{mp} = f_{source}$. Under this condition, Eq. (7) can be rearranged to express the magnetic field as a function of the top-gate voltage $V_{TG}$. In the case of the Dirac magnetoplasmon in monolayer graphene, this relation takes the form:

$$B^D = \sqrt{\left[4\pi^2 f_{Source}^2 - \frac{e^{3/2}v_F}{\hbar}\left(\frac{|V_{TG}|}{\pi C_{TG}}\right)^{1/2}\left(\frac{\pi}{2L_{TG}}(2N+1)\right)^2\right]\frac{\pi C_{TG}|V_{TG}|}{e^3}\left(\frac{\hbar}{v_f}\right)} \qquad (10)$$

And for the so-called Schrödinger-type magnetoplasmon in bilayer graphene:

$$B^S = \frac{m_*}{e}\sqrt{\left[4\pi^2 f_{Source}^2 - \left(\frac{e}{m_*}|V_{TG}|\right)\left(\frac{1\pi}{2L_{TG}}(2N+1)\right)^2\right]} \qquad (11)$$

Accordingly, the magnetic field where magnetoplasmon resonance arise can be represented as a function of the top-gate voltage. **Figure 3** presents a comparative analysis between Dirac-type (monolayer) and Schrödinger-type (bilayer) systems, highlighting the impact of the effective mass on the magnetoplasmon dispersion.

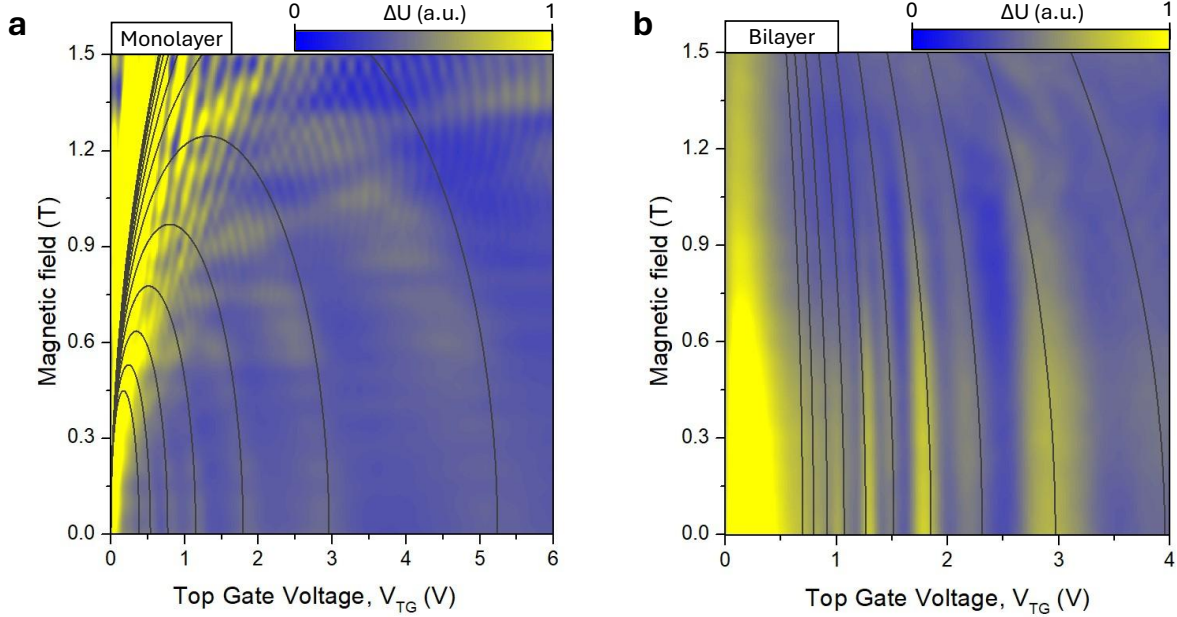

**Figure 3. Resonant magnetoplasmons in graphene TeraFETs.** Mapping of the photoresponse as a function of magnetic field and top-gate voltage under 3.5 THz illumination, measured at 1.7 K for monolayer graphene, **a**, and at 15K for bilayer graphene, **b**. The yellow color indicates those values with higher intensity measured in the THz photoresponse mapping while blue color indicates the lower values. Theoretical curves for the magnetoplasmon resonance positions are shown in gray lines in both mappings, calculated using Eq. (10) for the Dirac (monolayer, panel **a**) system and Eq. (11) for the Schrödinger (bilayer, panel **b**) system.

Theoretical resonance positions, obtained from equations (10) and (11), are superimposed on the color maps of the measured THz photoresponse as a function of magnetic field and top-gate voltage, under 3.5 THz illumination at 1.7 K and 15 K, for monolayer and bilayer graphene TeraFETs, respectively. In monolayer graphene (see **Figure 3a**), Shubnikov–de Haas oscillations become increasingly pronounced near the CNP as the magnetic field increases, thereby hindering experimental access to the low-density and low-field regime and progressively obscuring the magnetoplasmon resonances. Nevertheless, the resonance that can be resolved follows the expected $|V_{TG}|^{-1/4}$ dependence, consistent with the Dirac model. Moreover, the analysis of the resonance indicates a non-monotonic dependence of the resonance magnetic field ($B_{res}$) on carrier density, as predicted by theory. This behaviour indicates a crossover between two distinct dynamical regimes: at low carrier density, the density-dependent cyclotron response of the Dirac spectrum drives the resonance toward zero magnetic field, whereas at higher densities the collective plasmonic response, determined by carrier density and spatial confinement, becomes dominant. This non-monotonic behaviour is a hallmark of Dirac carriers, reflecting their density-dependent cyclotron response and distinguishing them from conventional parabolic-band systems. The maximum in $B_{res}(n)$ at intermediate densities reveals pronounced cyclotron–plasmon hybridization, with the resonance governed by the interplay of both modes.

In contrast to the monolayer graphene, measurements in bilayer graphene revealed universal conductance fluctuations (UCF)[37], becoming prominent at temperatures below 10K, preventing the clear resolution and observation of graphene magnetoplasmon resonances. However, to suppress these phase-coherence effects, measurements were intentionally performed at higher temperatures (15 K), which allows at this temperature the clear observation of the magnetoplasmons. A comparative dataset at 1.7 K and 15 K is included in ***Supplementary Information Note 7***, demonstrating that the suppression of UCF at higher temperature is required to clearly resolve the magnetoplasmon resonances. Therefore, **Figure 3b** shows the photoresponse mapping in bilayer graphene TeraFETs but at moderate temperatures of 15 K, modeled in the Schrödinger framework with a fixed effective mass of $m*/m_0 = 0.0026$[35]. The resonance exhibits a $|V_{TG}|^{-1/2}$ scaling, consistent with a Schrödinger-type description. In this case, the resonance tends toward the cyclotron frequency at low carrier density, as expected for systems characterized by an effective mass. For both monolayer and bilayer graphene, the agreement with theoretical predictions extends over a wide frequency range, from 1 THz to 3.5 THz, as shown in the *Supplementary Information Note 8*.

To the best of our knowledge, the only previous study closely related to the resonant Dyakonov–Shur plasma-wave model under a perpendicular magnetic field is that of Park et al.[36], who reported magnetoplasmon resonances up to 400 GHz in gated GaAs/AlGaAs mesa structures using on-chip THz time-domain spectroscopy. While their work represents an important milestone in conventional III–V 2DEG systems, our study advances the field by demonstrating clear signatures of cavity-confined magnetoplasmon resonances in a fundamentally different material platform across the THz range (up to 3.8 THz). Here we move beyond conventional III–V 2DEGs with parabolic dispersion to establish the robustness and universality of the Dyakonov–Shur magneto-resonant regime in monolayer graphene, a unique platform with a linear, gapless Dirac band structure. In parallel, bilayer graphene offers a complementary parabolic system that, unlike GaAs-based heterostructures, features a gate-tunable bandgap (via displacement field), ambipolar operation with the gate bias, and low-disorder hBN encapsulation enabling several long-lived cavity plasmons. These attributes provide an alternative route to resonant plasmonics not accessible in conventional semiconductors and allow a direct comparison with the gapless, linear monolayer case. The observation of magnetoplasmon resonances in both mono- and bilayer graphene highlights not only the adaptability of the Dyakonov–Shur mechanism to emerging 2D materials, but also its relevance in regimes far from those traditionally explored in III–V heterostructures. Overall, we show a good agreement between experiment and theory in both monolayer and bilayer graphene systems. As shown in **Figure 2** and **3** in the main text and also additional data presented in the ***Supporting Information Note 8***, minor deviations can be observed at B = 0 in both materials, despite the overall good agreement provided by the parabolic-band model for the bilayer and the Dirac model for the monolayer. These slightly low-field mismatches can reasonably be ascribed to uncertainties in electron density calibration[38], global parameters such as the Fermi velocity[39,40] or oxide capacitance[41], as well as possible fringing effects[42]. Beyond this baseline discrepancy, the monolayer exhibits an additional deviation at finite magnetic field. Although the initial field dependence follows perfectly theoretical prediction, the experimental curves consistently lie below the calculated ones near the inflection point, corresponding to the maximum of the resonance magnetic field. The origin of this additional shift remains uncertain. Non-local electrodynamics could, in principle, contribute to the effect[43–45], since incorporating non-local terms in the dispersion lowers the resonance maximum in qualitative agreement with experiment. However, this interpretation should be regarded as tentative, because non-locality would also alter the dispersion at low fields, where theory and experiment otherwise agree well in some of the spectra. However, this interpretation should be regarded as tentative, because non-locality would also alter the dispersion at low fields, where theory and experiment otherwise agree well in some of the spectra. In addition, electron–electron interactions may play a significant role. These interactions are known to renormalize the Fermi velocity, particularly at low carrier concentrations[46,47] leading to an enhancement of $v_F$ and, consequently, to a downward shift of the resonant magnetic-field maximum. Finally, edge magnetoplasmons and their possible coupling to bulk modes[26] could also contribute to the observed resonance shift. Such combined effects involving geometric factors, non-local corrections, and many-body interactions, need further investigation to fully elucidate the origin of the deviations observed in the magnetoplasmon dispersion of graphene.

**Conclusions**

In summary, we have investigated resonant Dyakonov–Shur magnetoplasmons in high-mobility mono- and bilayer graphene TeraFETs subjected to THz radiation. In absence of magnetic field, both systems present univocal fingerprints of plasmon-assisted resonant THz photodetection, providing a direct comparison between two unique systems with distinct plasma-wave dispersion in a wide frequency operation range (2.5 – 3.8 THz). For a finite magnetic field, perpendicularly applied to the systems, the observed resonant magnetoplasmon modes are due to the coupling of standing plasma-waves under the gate with the cyclotron motion of carriers. In bilayer graphene, the magnetoplasmon dispersion is well captured by a Schrödinger-type model with a constant effective mass, whereas in monolayer graphene it follows the Dirac picture with a density-dependent cyclotron response. This leads to a non-monotonic evolution of the Dirac magnetoplasmon dispersion with carrier density, including an inflection point where the coupling between the plasmon and the cyclotron resonance is strongest. Overall, the theoretical framework reproduces the experimental dispersion well across a broad range of parameters such as gate voltage, magnetic field, and frequency. These findings establish graphene TeraFETs as a robust platform for exploring magnetoplasmons and for advancing future reconfigurable terahertz plasmonic sensing technologies.

## ASSOCIATED CONTENT

The following files are available free of charge:

- Supporting Information (PDF): Additional fabrication details, Plasmon lifetime analysis and additional data related with temperature- and frequency-dependent measurements.

## ACKNOWLEDGMENT

This work was supported by the Terahertz Occitanie Platform, by the French "Agence Nationale pour la Recherche" for TEASER project (ANR-24-CE24-4830), by the "France 2030 program" through the Equipex+ HYBAT project (ANR-21-ESRE-0026), and by the European Union through the Flag-Era JTC 2019 DeMeGras project (VR2019-00404, DFG No. GA 501/16-1, and ANR-19-GRF1-0006). This work was also supported by the Ministry of Science and Innovation under the grants PID2021-128154NAI00, CNS2024-154588, PDC2023-145856-I00 and PID2021-126483OB-I00. Financial support of the Department of Education of the Junta de Castilla y León and FEDER Funds is also gratefully acknowledged (Escalera de Excelencia CLU-2023-1-02). K.W. and T.T. acknowledge support from the JSPS KAKENHI (Grant Numbers 21H05233 and 23H02052), the CREST (JPMJCR24A5), JST and World Premier International Research Center Initiative (WPI), MEXT, Japan. J.A.D-N and J.M.C. acknowledge the support of the Grants RYC2023-044965-I and RYC2019-028443-I funded by MICIU/AEI/10.13039/501100011033 and by "ESF+" and the Research Grant number SA103P23 funded by Junta de Castilla y León and FEDER. J.M.C. also acknowledges financial support of the European Research Council (ERC) under Starting grant ID 101039754, CHIROTRONICS, funded by the European Union. Views and opinions expressed are, however, those of the author(s) only and do not necessarily reflect those of the European Union or the European Research Council. Neither the European Union nor the granting authority can be held responsible for them.

**REFERENCES**

(1) Basov, D. N.; Fogler, M. M.; García De Abajo, F. J. Polaritons in van Der Waals Materials. *Science (1979)* **2016**, *354* (6309). https://doi.org/10.1126/SCIENCE.AAG1992.

(2) Koppens, F. H. L.; Chang, D. E.; García De Abajo, F. J. Graphene Plasmonics: A Platform for Strong Light–Matter Interactions. *Nano Lett* **2011**, *11* (8), 3370–3377. https://doi.org/10.1021/NL201771H.

(3) De Abajo, F. J. G. Graphene Plasmonics: Challenges and Opportunities. *ACS Photonics* **2014**, *1* (3), 133–152. https://doi.org/10.1021/PH400147Y.

(4) Markelz, A. G.; Mittleman, D. M. Perspective on Terahertz Applications in Bioscience and Biotechnology. *ACS Photonics* **2022**, *9* (4), 1117–1126. https://doi.org/10.1021/acsphotonics.2c00228.

(5) Low, T.; Avouris, P. Graphene Plasmonics for Terahertz to Mid-Infrared Applications. *ACS Nano* **2014**, *8* (2), 1086–1101. https://doi.org/10.1021/nn406627u.

(6) Dyakonov, M.; Shur, M. Detection, Mixing, and Frequency Multiplication of Terahertz Radiation by Two-Dimensional Electronic Fluid. *IEEE Trans Electron Devices* **1996**, *43* (3), 380–387. https://doi.org/10.1109/16.485650.

(7) Dyakonov, M.; Shur, M. Shallow Water Analogy for a Ballistic Field Effect Transistor: New Mechanism of Plasma Wave Generation by Dc Current. *Phys Rev Lett* **1993**, *71* (15), 2465–2468. https://doi.org/10.1103/PhysRevLett.71.2465.

(8) Knap, W.; Teppe, F.; Meziani, Y.; Dyakonova, N.; Lusakowski, J.; Boeuf, F.; Skotnicki, T.; Maude, D.; Rumyantsev, S.; Shur, M. S. Plasma Wave Detection of Sub-Terahertz and Terahertz Radiation by Silicon Field-Effect Transistors. *Appl Phys Lett* **2004**, *85* (4), 675–677. https://doi.org/10.1063/1.1775034.

(9) Knap, W.; Kachorovskii, V.; Deng, Y.; Rumyantsev, S.; Lü, J. Q.; Gaska, R.; Shur, M. S.; Simin, G.; Hu, X.; Khan, M. A.; Saylor, C. A.; Brunel, L. C. Nonresonant Detection of Terahertz Radiation in Field Effect Transistors. *J Appl Phys* **2002**, *91* (11), 9346–9353. https://doi.org/10.1063/1.1468257.

(10) El Fatimy, A.; Boubanga Tombet, S.; Teppe, F.; Knap, W.; Veksler, D. B.; Rumyantsev, S.; Shur, M. S.; Pala, N.; Gaska, R.; Fareed, Q.; Hu, X.; Seliuta, D.; Valusis, G.; Gaquiere, C.; Theron, D.; Cappy, A. Terahertz Detection by GaN/AlGaN Transistors. *Electron Lett* **2006**, *42* (23), 1342–1344. https://doi.org/10.1049/EL:20062452.

(11) Kadykov, A. M.; Teppe, F.; Consejo, C.; Viti, L.; Vitiello, M. S.; Krishtopenko, S. S.; Ruffenach, S.; Morozov, S. V.; Marcinkiewicz, M.; Desrat, W.; Dyakonova, N.; Knap, W.; Gavrilenko, V. I.; Mikhailov, N. N.; Dvoretsky, S. A. Terahertz Detection of Magnetic

Field-Driven Topological Phase Transition in HgTe-Based Transistors. *Appl Phys Lett* **2015**, *107* (15), 56. https://doi.org/10.1063/1.4932943.

(12) Delgado-Notario, J. A.; Knap, W.; Clericò, V.; Salvador-Sánchez, J.; Calvo-Gallego, J.; Taniguchi, T.; Watanabe, K.; Otsuji, T.; Popov, V. V; Fateev, D. V; Diez, E.; Velázquez-Pérez, J. E.; Meziani, Y. M. Enhanced Terahertz Detection of Multigate Graphene Nanostructures. *Nanophotonics* **2022**, *11* (3), 519–529. https://doi.org/doi:10.1515/nanoph-2021-0573.

(13) Viti, L.; Hu, J.; Coquillat, D.; Politano, A.; Knap, W.; Vitiello, M. S. Efficient Terahertz Detection in Black-Phosphorus Nano-Transistors with Selective and Controllable Plasma-Wave, Bolometric and Thermoelectric Response. *Sci Rep* **2016**, *6* (1), 20474. https://doi.org/10.1038/srep20474.

(14) Vicarelli, L.; Vitiello, M. S.; Coquillat, D.; Lombardo, A.; Ferrari, A. C.; Knap, W.; Polini, M.; Pellegrini, V.; Tredicucci, A. Graphene Field-Effect Transistors as Room-Temperature Terahertz Detectors. *Nat Mater* **2012**, *11* (10), 865–871. https://doi.org/10.1038/nmat3417.

(15) Zak, A.; Andersson, M. A.; Bauer, M.; Matukas, J.; Lisauskas, A.; Roskos, H. G.; Stake, J. Antenna-Integrated 0.6 THz FET Direct Detectors Based on CVD Graphene. *Nano Lett* **2014**, *14* (10), 5834–5838. https://doi.org/10.1021/nl5027309.

(16) Delgado-Notario, J. A.; Power, S. R.; Knap, W.; Pino, M.; Cheng, J.; Vaquero, D.; Taniguchi, T.; Watanabe, K.; Jesús, J.; Velázquez-Pérez, E.; Meziani, Y. M.; Alonso-González, P.; Caridad, J. M. Unveiling the Miniband Structure of Graphene Moiré Superlattices via Gate-Dependent Terahertz Photocurrent Spectroscopy. *ACS Nano* **2025**. https://doi.org/10.1021/ACSNANO.5C05306.

(17) Bandurin, D. A.; Svintsov, D.; Gayduchenko, I.; Xu, S. G.; Principi, A.; Moskotin, M.; Tretyakov, I.; Yagodkin, D.; Zhukov, S.; Taniguchi, T.; Watanabe, K.; Grigorieva, I. V; Polini, M.; Goltsman, G. N.; Geim, A. K.; Fedorov, G. Resonant Terahertz Detection Using Graphene Plasmons. *Nat Commun* **2018**, *9* (1), 5392. https://doi.org/10.1038/s41467-018-07848-w.

(18) Caridad, J. M.; Castelló, Ó.; López Baptista, S. M.; Taniguchi, T.; Watanabe, K.; Roskos, H. G.; Delgado-Notario, J. A. Room-Temperature Plasmon-Assisted Resonant THz Detection in Single-Layer Graphene Transistors. *Nano Lett* **2024**, *24* (3), 935–942. https://doi.org/10.1021/acs.nanolett.3c04300.

(19) Man, L. F.; Xu, W.; Xiao, Y. M.; Wen, H.; Ding, L.; Van Duppen, B.; Peeters, F. M. Terahertz Magneto-Optical Properties of Graphene Hydrodynamic Electron Liquid. *Phys Rev B* **2021**, *104* (12), 125420. https://doi.org/10.1103/PhysRevB.104.125420.

(20) Kapralov, K.; Svintsov, D. Ballistic-to-Hydrodynamic Transition and Collective Modes for Two-Dimensional Electron Systems in Magnetic Field. *Phys Rev B* **2022**, *106* (11), 115415. https://doi.org/10.1103/PhysRevB.106.115415.

(21) Shen, Q.; Yan, J.; You, Y.; Li, S.; Shen, L. Terahertz Large-Area Unidirectional Surface Magnetoplasmon and Its Applications. *Sci Rep* **2023**, *13* (1), 1–10. https://doi.org/10.1038/S41598-023-49348-Y.

(22) Chandra, S.; Cozart, J.; Biswas, A.; Lee, S.; Chanda, D. Magnetoplasmons for Ultrasensitive Label-Free Biosensing. *ACS Photonics* **2021**, *8* (5), 1316–1323. https://doi.org/10.1021/ACSPHOTONICS.0C01646.

(23) Poumirol, J. M.; Yu, W.; Chen, X.; Berger, C.; De Heer, W. A.; Smith, M. L.; Ohta, T.; Pan, W.; Goerbig, M. O.; Smirnov, D.; Jiang, Z. Magnetoplasmons in Quasineutral Epitaxial Graphene Nanoribbons. *Phys Rev Lett* **2013**, *110* (24), 246803. https://doi.org/10.1103/PhysRevLett.110.246803.

(24) Crassee, I.; Orlita, M.; Potemski, M.; Walter, A. L.; Ostler, M.; Seyller, T.; Gaponenko, I.; Chen, J.; Kuzmenko, A. B. Intrinsic Terahertz Plasmons and Magnetoplasmons in Large Scale Monolayer Graphene. *Nano Lett* **2012**, *12* (5), 2470–2474. https://doi.org/10.1021/NL300572Y.

(25) Bartolomei, H.; Bisognin, R.; Kamata, H.; Berroir, J. M.; Bocquillon, E.; Ménard, G.; Plaçais, B.; Cavanna, A.; Gennser, U.; Jin, Y.; Degiovanni, P.; Mora, C.; Fève, G. Observation of Edge Magnetoplasmon Squeezing in a Quantum Hall Conductor. *Phys Rev Lett* **2023**, *130* (10), 106201. https://doi.org/10.1103/PhysRevLett.130.106201.

(26) Yan, H.; Li, Z.; Li, X.; Zhu, W.; Avouris, P.; Xia, F. Infrared Spectroscopy of Tunable Dirac Terahertz Magneto-Plasmons in Graphene. *Nano Lett* **2012**, *12* (7), 3766–3771. https://doi.org/10.1021/NL3016335.

(27) Eriksen, M. H.; Deop-Ruano, J. R.; Cox, J. D.; Manjavacas, A. Chiral Light–Matter Interactions with Thermal Magnetoplasmons in Graphene Nanodisks. *Nano Lett* **2024**, *25* (1), 313–320. https://doi.org/10.1021/ACS.NANOLETT.4C05056.

(28) Papasimakis, N.; Thongrattanasiri, S.; Zheludev, N. I.; García De Abajo, F. J. The Magnetic Response of Graphene Split-Ring Metamaterials. *Light Sci Appl* **2013**, *2* (JULY), e78–e78. https://doi.org/10.1038/LSA.2013.34.

(29) Frigerio, E.; Rebora, G.; Ruelle, M.; Souquet-Basiège, H.; Jin, Y.; Gennser, U.; Cavanna, A.; Plaçais, B.; Baudin, E.; Berroir, J. M.; Safi, I.; Degiovanni, P.; Fève, G.; Ménard, G. C. Gate Tunable Edge Magnetoplasmon Resonators. *Commun Phys* **2024**, *7* (1), 1–8. https://doi.org/10.1038/S42005-024-01803-6.

(30) Hiyama, N.; Hashisaka, M.; Fujisawa, T. An Edge-Magnetoplasmon Mach-Zehnder Interferometer. *Appl Phys Lett* **2015**, *107* (14), 143101. https://doi.org/10.1063/1.4932111.

(31) Zhao, W.; Wang, S.; Chen, S.; Zhang, Z.; Watanabe, K.; Taniguchi, T.; Zettl, A.; Wang, F. Observation of Hydrodynamic Plasmons and Energy Waves in Graphene. *Nature* **2023**, *614* (7949), 688–693. https://doi.org/10.1038/S41586-022-05619-8.

(32) Winstanley, B.; Schomerus, H.; Principi, A. Corbino Field-Effect Transistors in a Magnetic Field: Highly Tunable Photodetectors. *Phys Rev B* **2021**, *104* (16), 165406. https://doi.org/10.1103/PhysRevB.104.165406.

(33) Tomadin, A.; Polini, M. Theory of the Plasma-Wave Photoresponse of a Gated Graphene Sheet. *Phys Rev B* **2013**, *88* (20), 205426. https://doi.org/10.1103/PhysRevB.88.205426.

(34) Chae, J.; Jung, S.; Young, A. F.; Dean, C. R.; Wang, L.; Gao, Y.; Watanabe, K.; Taniguchi, T.; Hone, J.; Shepard, K. L.; Kim, P.; Zhitenev, N. B.; Stroscio, J. A. Renormalization of the Graphene Dispersion Velocity Determined from Scanning Tunneling Spectroscopy. *Phys Rev Lett* **2012**, *109* (11), 116802. https://doi.org/10.1103/PhysRevLett.109.116802.

(35) Zou, K.; Hong, X.; Zhu, J. Effective Mass of Electrons and Holes in Bilayer Graphene: Electron-Hole Asymmetry and Electron-Electron Interaction. *Phys Rev B* **2011**, *84* (8), 085408. https://doi.org/10.1103/PhysRevB.84.085408.

(36) Cunningham, J. E.; Linfield, E. H.; Wood, C. D.; Li, L. H.; Wu, J.; Park, S. J.; Davies, A. G.; Zonetti, S.; Sydoruk, O.; Parker-Jervis, R. S. Terahertz Magnetoplasmon Resonances in Coupled Cavities Formed in a Gated Two-Dimensional Electron Gas. *Optics Express, Vol. 29, Issue 9, pp. 12958-12966* **2021**, *29* (9), 12958–12966. https://doi.org/10.1364/OE.414178.

(37) Kharitonov, M. Yu.; Efetov, K. B. Mesoscopic Conductance Fluctuations in Graphene Samples. *Phys Rev B* **2008**, *78* (3), 033404. https://doi.org/10.1103/PhysRevB.78.033404.

(38) Xia, F.; Mueller, T.; Lin, Y. M.; Valdes-Garcia, A.; Avouris, P. Ultrafast Graphene Photodetector. *Nat Nanotechnol* **2009**, *4* (12), 839–843. https://doi.org/10.1038/NNANO.2009.292.

(39) Elias, D. C.; Gorbachev, R. V.; Mayorov, A. S.; Morozov, S. V.; Zhukov, A. A.; Blake, P.; Ponomarenko, L. A.; Grigorieva, I. V.; Novoselov, K. S.; Guinea, F.; Geim, A. K. Dirac Cones Reshaped by Interaction Effects in Suspended Graphene. *Nat Phys* **2011**, *7* (9), 701–704. https://doi.org/10.1038/NPHYS2049.

(40) Whelan, P. R.; Shen, Q.; Zhou, B.; Serrano, I. G.; Kamalakar, M. V.; MacKenzie, D. M. A.; Ji, J.; Huang, D.; Shi, H.; Luo, D.; Wang, M.; Ruoff, R. S.; Jauho, A. P.; Jepsen, P. U.; Bøggild, P.; Caridad, J. M. Fermi Velocity Renormalization in Graphene Probed by Terahertz Time-Domain Spectroscopy. *2d Mater* **2020**, *7* (3), 035009. https://doi.org/10.1088/2053-1583/AB81B0.

(41) Fei, Z.; Rodin, A. S.; Andreev, G. O.; Bao, W.; McLeod, A. S.; Wagner, M.; Zhang, L. M.; Zhao, Z.; Thiemens, M.; Dominguez, G.; Fogler, M. M.; Castro Neto, A. H.; Lau, C. N.;

Keilmann, F.; Basov, D. N. Gate-Tuning of Graphene Plasmons Revealed by Infrared Nano-Imaging. *Nature* **2012**, *486* (7405), 82–85. https://doi.org/10.1038/NATURE11253.

(42) Nishimura, T.; Magome, N.; Khmyrova, I.; Suemitsu, T.; Knap, W.; Otsuji, T. Analysis of Fringing Effect on Resonant Plasma Frequency in Plasma Wave Devices. *Jpn J Appl Phys* **2009**, *48* (4 PART 2), 04C096. https://doi.org/10.1143/JJAP.48.04C096.

(43) Pack, J.; Russell, B. J.; Kapoor, Y.; Balgley, J.; Ahlers, J.; Taniguchi, T.; Watanabe, K.; Henriksen, E. A. Broken Symmetries and Kohn's Theorem in Graphene Cyclotron Resonance. *Phys Rev X* **2020**, *10* (4), 041006. https://doi.org/10.1103/PhysRevX.10.041006.

(44) Kiselev, E. I.; Schmalian, J. Nonlocal Hydrodynamic Transport and Collective Excitations in Dirac Fluids. *Phys Rev B* **2020**, *102* (24), 245434. https://doi.org/10.1103/PhysRevB.102.245434.

(45) Rodriguez-Lopez, P.; Wang, J. S.; Antezza, M. Electric Conductivity in Graphene: Kubo Model versus a Nonlocal Quantum Field Theory Model. *Phys Rev B* **2025**, *111* (11), 115428. https://doi.org/10.1103/PhysRevB.111.115428.

(46) Yu, G. L.; Jalil, R.; Belle, B.; Mayorov, A. S.; Blake, P.; Schedin, F.; Morozov, S. V.; Ponomarenko, L. A.; Chiappini, F.; Wiedmann, S.; Zeitler, U.; Katsnelson, M. I.; Geim, A. K.; Novoselov, K. S.; Elias, D. C. Interaction Phenomena in Graphene Seen through Quantum Capacitance. *Proc Natl Acad Sci U S A* **2013**, *110* (9), 3282–3286. https://doi.org/10.1073/PNAS.1300599110.

(47) Faugeras, C.; Berciaud, S.; Leszczynski, P.; Henni, Y.; Nogajewski, K.; Orlita, M.; Taniguchi, T.; Watanabe, K.; Forsythe, C.; Kim, P.; Jalil, R.; Geim, A. K.; Basko, D. M.; Potemski, M. Landau Level Spectroscopy of Electron-Electron Interactions in Graphene. *Phys Rev Lett* **2015**, *114* (12), 126804. https://doi.org/10.1103/PHYSREVLETT.114.126804.

# Supporting Information for

# Resonant Dyakonov–Shur Magnetoplasmons in Graphene Terahertz Photodetectors

*Juan A. Delgado-Notario[1,2][⊥]\*, Cedric Bray[3][⊥], Elsa Perez-Martin[3], Ben Benhamou-Bui[3], Namrata Saha[1,2], Sahil Parvez[1,2], Christophe.Consejo[3], Guillaume Sigu[3], Salah Benlemqwanssa[3], Laurent Bonnet[3], Takashi Taniguchi[4], Kenji Watanabe[5], José M. Caridad[1,2], Sergey S. Krishtopenko[3], Yahya M. Meziani[1], Benoit Jouault[3], Jérémie Torres[3], Sandra Ruffenach[3] and Frédéric Teppe[3]\**

[1] Departamento de Física Aplicada, Universidad de Salamanca, 37008 Salamanca, Spain

[2] Unidad de Excelencia en Luz y Materia Estructurada (LUMES), Universidad de Salamanca, Salamanca 37008, Spain

[3] L2C (UMR 5221), Université de Montpellier, CNRS, Montpellier, France

[4] *Research Center for Materials Nanoarchitectonics, National Institute for Materials Science, 1-1 Namiki, Tsukuba 305-0044, Japan*

[5] Research Center for Electronic and Optical Materials, National Institute for Materials Science, 1-1 Namiki, Tsukuba 305-0044, Japan

\* juanandn@usal.es, frederic.teppe@umontpellier.fr

[⊥] These authors contributed equally

**Supporting information Note 1: Device fabrication**

All devices were fabricated on highly doped Si substrates (serving as a global back gate) coated with 300 nm thermally grown $SiO_2$. The fabrication flow for mono- and bilayer graphene TeraFETs was identical. First, graphite and hBN crystals were exfoliated using adhesive tape and assembled into hBN/graphene/hBN heterostructures (either mono- or bilayer graphene) by a van der Waals dry-stacking technique[1,2]. For this step, we intentionally selected elongated and relatively narrow (4 to 6 µm) graphene flakes of nearly constant width (**Figure S1a**) together with wider hBN crystals (**Figure S1b**). This selection facilitates defining the top-gate metallic contact and the channel width without chemical patterning of the graphene heterostructure (**Figure S1c**) and thereby preserving the high intrinsic quality of the selected graphene flakes[3].

The resulting stacks (based on monolayer or bilayer graphene) were then patterned by electron-beam lithography (EBL) to define and open the contact windows to the graphene, electrodes and antenna lobes, using PMMA (6% in chlorobenzene) as the resist. The heterostructures were subsequently dry-etched in an ICP–RIE (Oxford PlasmaPro Cobra 100) with $SF_6$ (40 sccm, 6 mTorr) at 75 W and 10 °C, followed by e-beam evaporation of Cr/Au (3.5/55 nm) to create the metal contacts/electrodes. Finally, a second EBL step was used to define the top gate, followed by e-beam evaporation of Cr/Au (5/45 nm).

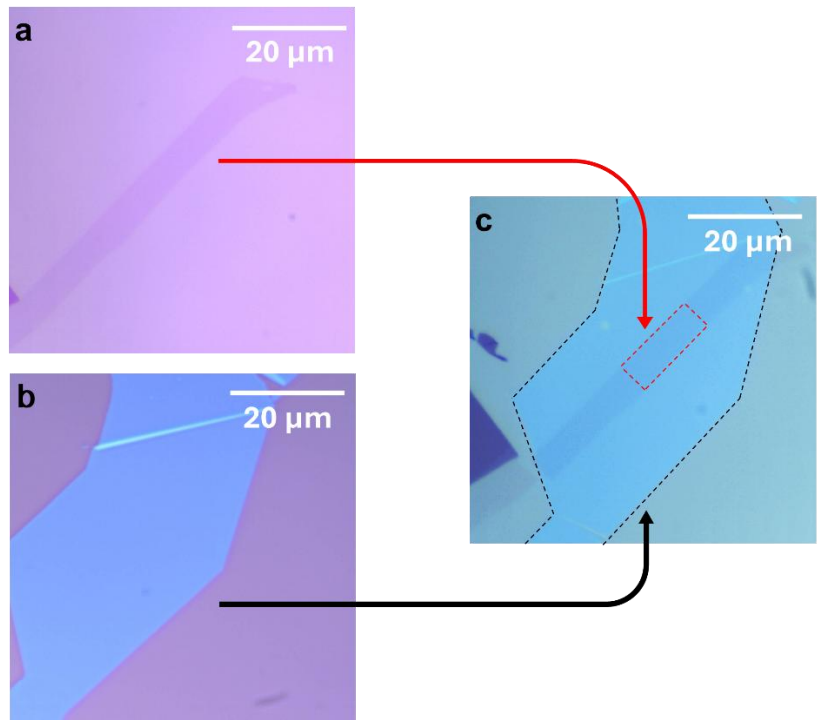

**Figure S1. Graphene based heterostructures.** Optical image of the typical exfoliated flakes of **a,** Graphene, **b,** hBN and **c,** the fabricated half-heterostructure where the dashed red lines highlight the graphene flake area with constnt width selected for the device fabrication. The hBN flake (highlighted with dashed black lines) was intentionally selected to fully cover the graphene flake.

**Supporting information Note 2: Broadband operation in Bilayer graphene TeraFETs**

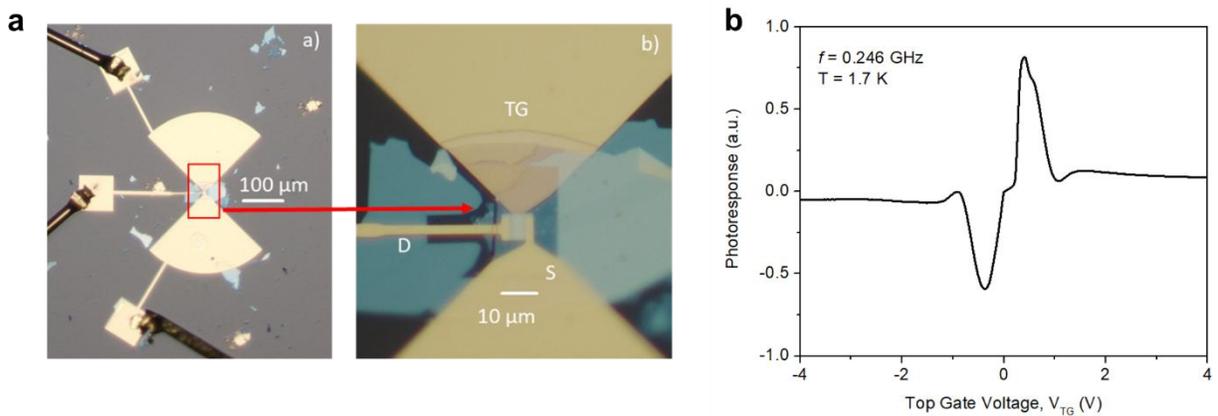

**Figure S2. Bilayer Graphene-based TeraFET photodetector. a**, Optical microscope image of the bilayer graphene TeraFET, featuring a bow-tie antenna integrated. The right panel provides a zoomed-in view of the device, with Source, Top gate, and Drain contacts labeled. **b**, THz photoresponse as a function of the top gate voltage, $V_{TG}$, measured for the bilayer graphene TeraFET operating in the non-resonant or broadband regime at 0.246 THz for a temperature of 1.7 K and zero magnetic field.

**Supporting information Note 3: Determination of the top-gate capacitance per unit area**

The top-gate capacitance per unit area was calculated by the formula:

$$C_{TG} = \frac{\varepsilon_{hBN}\varepsilon_0}{t_{hBN}} \tag{S1}$$

Where $\varepsilon_{hBN}$ is the relative permittivity of hBN, $\varepsilon_0$ is the vacuum permittivity and $t_{hBN}$ the hBN thickness is approximately close to 28 nm in our monolayer and bilayer graphene-based devices. Using this value, a total gate capacitance of 0.00112 F.m$^{-2}$ can be found. In addition, the hBN gate capacitance per unit area can be determinate by the Shubnikov–de Haas oscillations[4]. The effective capacitance between the top gate and the graphene-based channel is given by:

$$C_{eff} = \frac{C_q}{C_q + C_{TG} + C_{BG}} \times C_{TG} \tag{S2}$$

where $C_q$ is the quantum capacitance and $C_{TG}$ and $C_{BG}$ are the capacitance of top gate and back gate respectively. To a first approximation, we can simplify equation S2 as $C_{eff} \sim C_{TG}$. This is confirmed by the periodicity of the Shubnikov–de Haas oscillations. As such, we can determinate the value of the hBN gate capacitance per unit area by the equation:

$$\Delta N = \frac{g_v eB}{\pi\hbar} = \frac{C_{TG}\Delta V_{TG}}{e} \tag{S3}$$

with $\Delta N$ the maximum number of electrons per unit area that each Landau level can accommodate, $g_v$ the valley degeneracy in graphene and $\Delta V_{TG}$ the increment of gate voltage necessary to create $\Delta N$ at the value of magnetic field, $B$, and equals the spacing between adjacent oscillation peaks. For example, in our case we obtained a value of 0.00101 F.m$^{-2}$ for the monolayer graphene. The slight discrepancy of a few percent below the theoretical value $C_{TG}$ of  is easily explained if one takes into account the full equation S2, including the prefactor $\frac{C_q}{C_q + C_{TG} + C_{BG}}$.

**Supporting information Note 4: Plasma wave resonant regime**

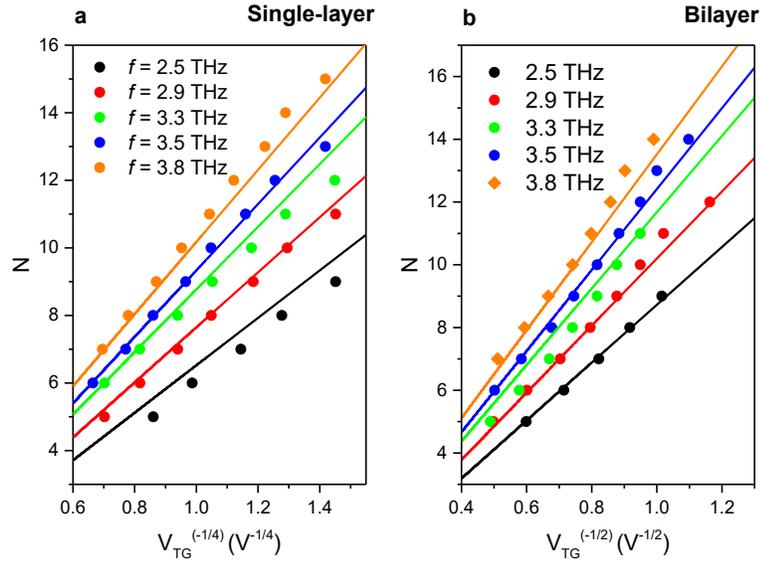

**Figure S3. Plasmon dispersion.** Resonant mode number of plasmons as a function of $|V_{TG}|^{-1/4}$ for monolayer device, **a**, and as function of $|V_{TG}|^{-1/2}$ for bilayer device, **b**, for illumination frequencies at 2.5, 2.9, 3.3, 3.5 and 3.8 THz (dot symbols). The observed trends show good agreement with the Dyakonov–Shur theory adapted to monolayer graphene and bilayer graphene (solid lines).

**Supporting information Note 5: Plasmon lifetime in monolayer and bilayer graphene**

The plasmon lifetime, $\tau_p$, can be related to the full width at half maximum (FWHM) of each resonant peak, $\delta$, observed in the photoresponse spectra via[5,6]:

$$\frac{V_G^{-1/4}}{\delta} = \omega\tau_p, \text{ for monolayer graphene} \qquad (S4)$$

$$\frac{V_G^{-1/2}}{\delta} = \omega\tau_p, \text{ for bilayer graphene} \qquad (S5)$$

where $\omega$ is the angular resonance frequency. To extract $\tau_p$, we perform simultaneous (global) Lorentzian fits of the set of resonant peaks observed in the THz photoresponse measurements for both monolayer and bilayer TeraFETs (**Figures S4 and S5**), according to the scaling across modes presented in **Equations S4** and **S5** respectively. The resulting lifetimes range from 0.4 to 1.7 ps in both systems. Consistently with previous works[5,6], these values are slightly longer than the transport time values extracted from mobility, showing that momentum relaxation alone cannot account for the observed broadening. This points to the presence of additional dissipation mechanisms, such as plasmon leakage into contacts[7], oblique plasma modes[8], and electronic viscosity[9]. Nevertheless, electronic viscosity is expected to lower the plasmon lifetime near the charge neutrality point, which does not match the behaviour observed in **Figures S4 and S5**, making it unlikely to play a major role here (we consistently observe a light enhancement of plasmon lifetimes as the charge-neutrality point is approached, which coincides with the regime where higher-order resonances are resolved.). Although hydrodynamic effects can be ruled out as the origin of the observed broadening, temperature-dependent studies would still be highly valuable to investigate in more detail the influence of electron–electron interactions. In monolayer graphene, however, our measurements showed no appreciable variation of the plasmon lifetime between 1.7 K and 15 K (see **Figure S4**). Moreover, modifying the channel geometry, as shown for example in Ref.[8] and Ref.[10], offers a promising route to suppress oblique plasma modes. Such suppression has already been demonstrated to reduce the resonance linewidth in multichannel HEMTs, and applying a similar strategy to graphene could help separate extrinsic effects from the intrinsic plasmon damping mechanisms. Finally, we have observed how the uncertainty increases for higher-order modes (i.e., at lower top-gate voltages), reflecting their reduced signal-to-noise after removing the non-resonant contribution.

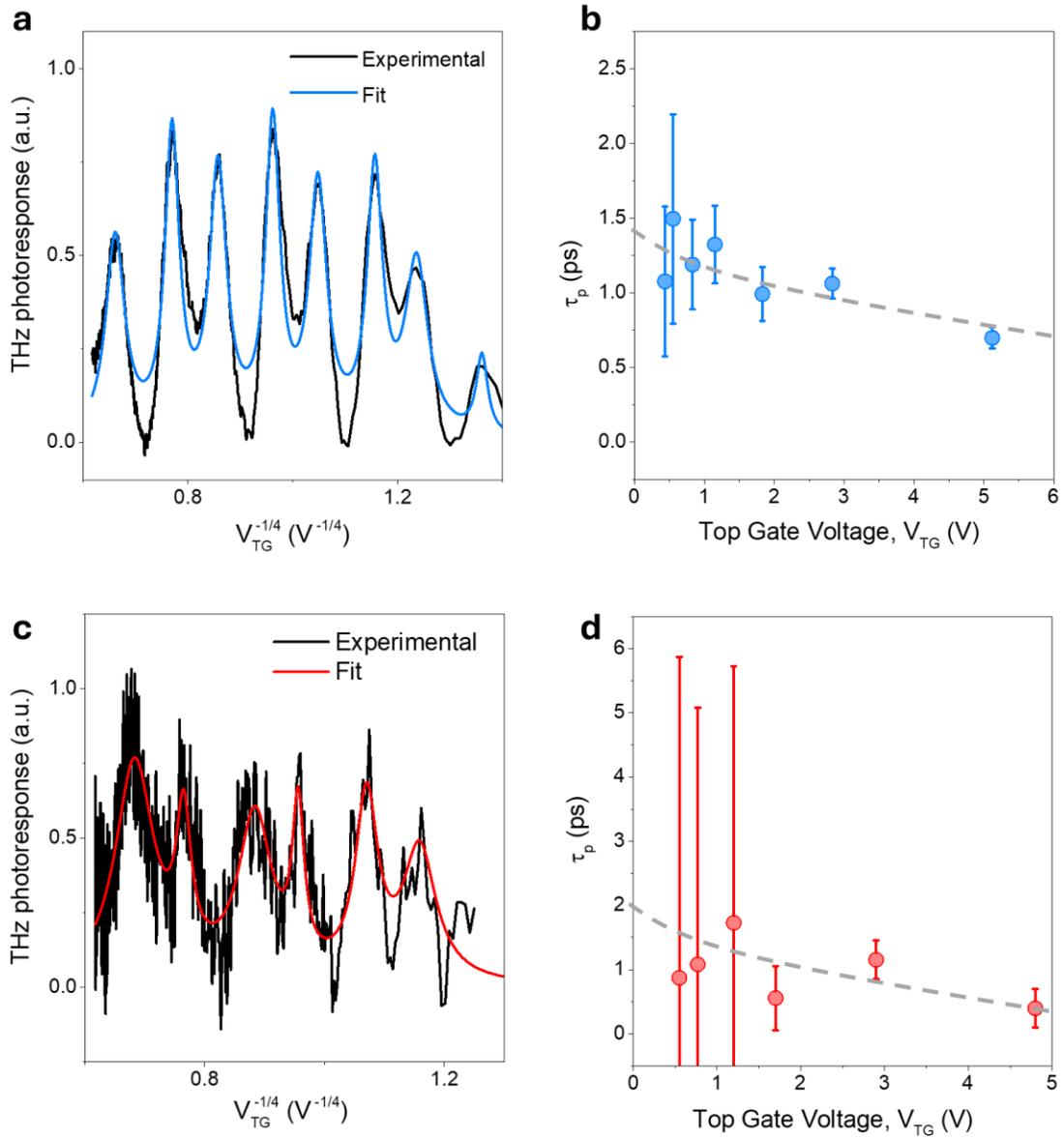

**Figure S4. Plasmon lifetime in monolayer graphene**. **a**, Experimental THz photoresponse for the monolayer graphene TeraFET as a function of $V_{TG}^{-1/4}$ (black line) upon removing the smooth non-resonant background at 1.7K for an excitation frequency of 3.5 THz. Solid blue corresponds to the global Lorentzian fit of the experimental data. **b**, Plasmon lifetime, $\tau_p = \frac{V_G^{-1/4}}{\delta\omega}$, as a function of the top gate voltage calculated for each mode by using the full widths at half maximum values, δ, extracted from the multi-Lorentzian fit displayed in panel **a**. Dashed grey line has been included to guide the eye. Panels **c** and **d** show the same measurements as **a** and **b**, respectively, but at 15K.

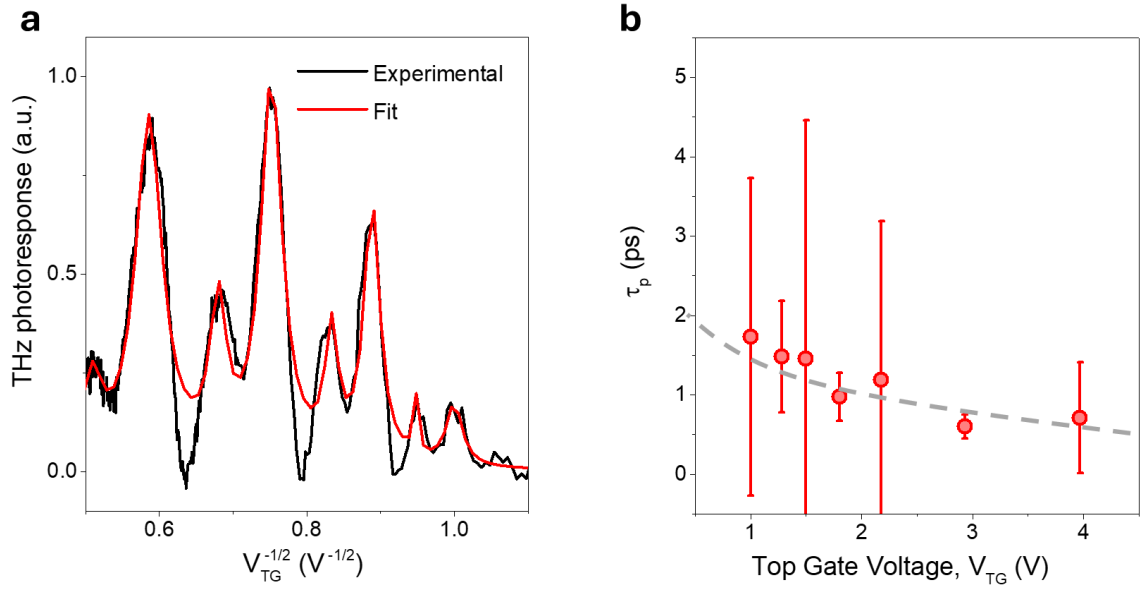

**Figure S5. Plasmon lifetime in bilayer graphene**. **a**, Experimental THz photoresponse in the bilayer graphene as a function of $V_{TG}^{-1/2}$ (black line) upon removing the smooth non-resonant background at 15K for an excitation frequency of 3.5 THz. Solid blue corresponds to the global Lorentzian fit of the experimental data. **b**, Plasmon lifetime, $\tau_p = \frac{V_G^{-1/2}}{\delta\omega}$, as a function of the top gate voltage calculated for each mode by using the full widths at half maximum values, δ, extracted from the multi-Lorentzian fit displayed in panel **a**. Dashed grey line has been included to guide the eye.

**Supporting information Note 6: Magnetoplasmons in monolayer graphene**

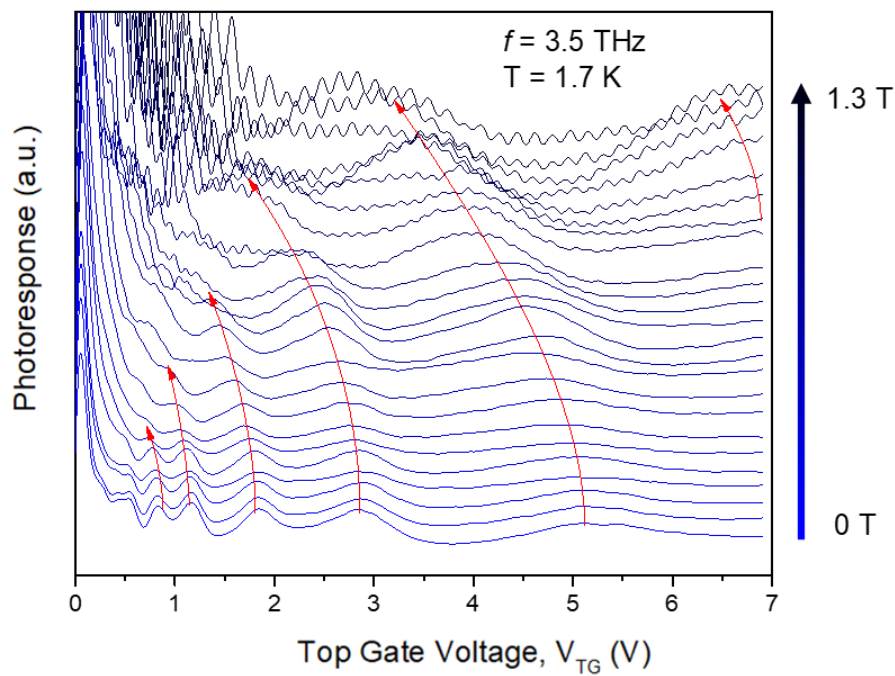

**Figure S6. Magnetoplasmons in monolayer graphene.** Gate-voltage dependence of the magnetoplasmon photoresponse measured at 1.7 K with 3.5 THz excitation under a perpendicular magnetic field up to 1.3 T. Curves have been vertical shifted for clarity. Close to the CNP, clear Shubnikov–de Haas oscillations arise at higher magnetic fields. Magnetoplasmon-related peaks (indicated by red arrows) shift progressively toward lower carrier densities (i.e., lower $V_{TG}$) as the magnetic field increases. This shift is more pronounced for lower-order modes.

**Supporting information Note 7: Universal conductance fluctuations in bilayer graphene**

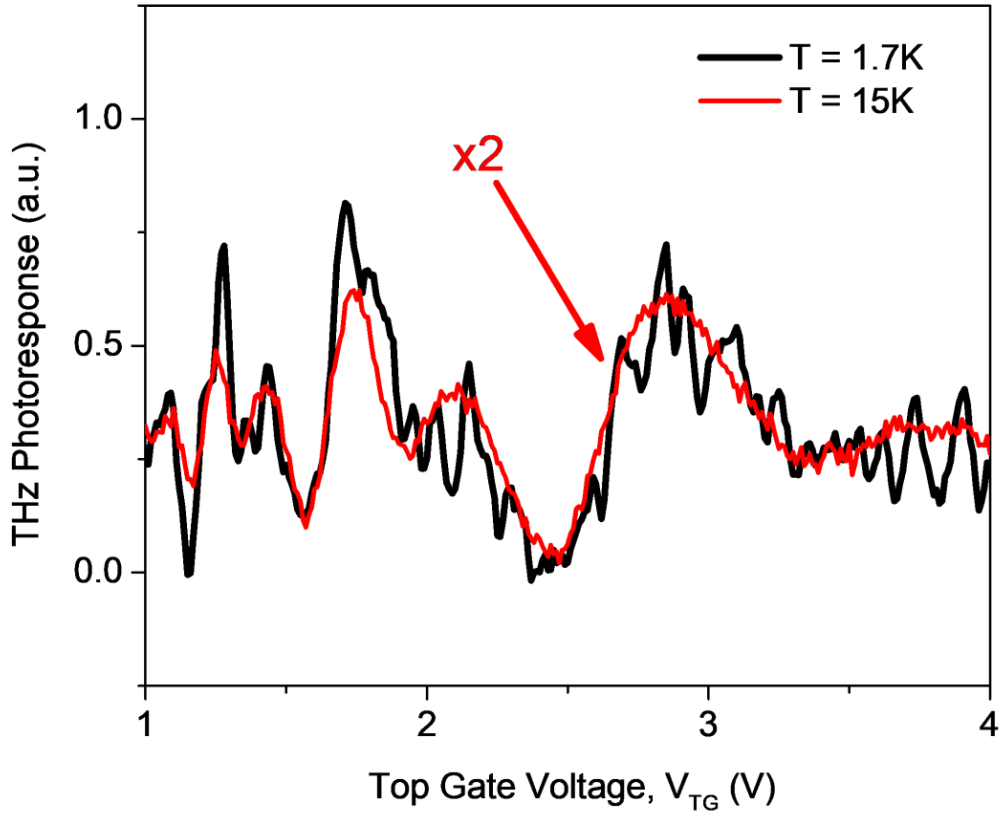

**Figure S7. Temperature dependence photoresponse in bilayer graphene**. Photoresponse as a function of the top gate voltage in the bilayer TeraFET for an illumination of 3.5 THz at 0.75T for a temperature of 1.7K (black curve) and 15K (red curve). The signal of photoresponse for the measurement at 15K has been multiplied by a factor x2 for clarity. The fine features observed at 1.7K arises from mesoscopic conductance fluctuations rather than noise. These fluctuations are thermally suppressed at 15K, making the plasmonic resonances visible.

**Supporting information Note 8: Additional frequency dependence data**

Observation of magnetoplasmons have been done on monolayer and bilayer graphene devices for frequencies between 2.5 THz and 3.8 THz. For the case of monolayer graphene TeraFETs, the temperature during the experimental measurement was set to 1.7K. In contrast, for the case of bilayer graphene TeraFETs, the operation temperature during the measurements was set to 15K to suppress UFCs.

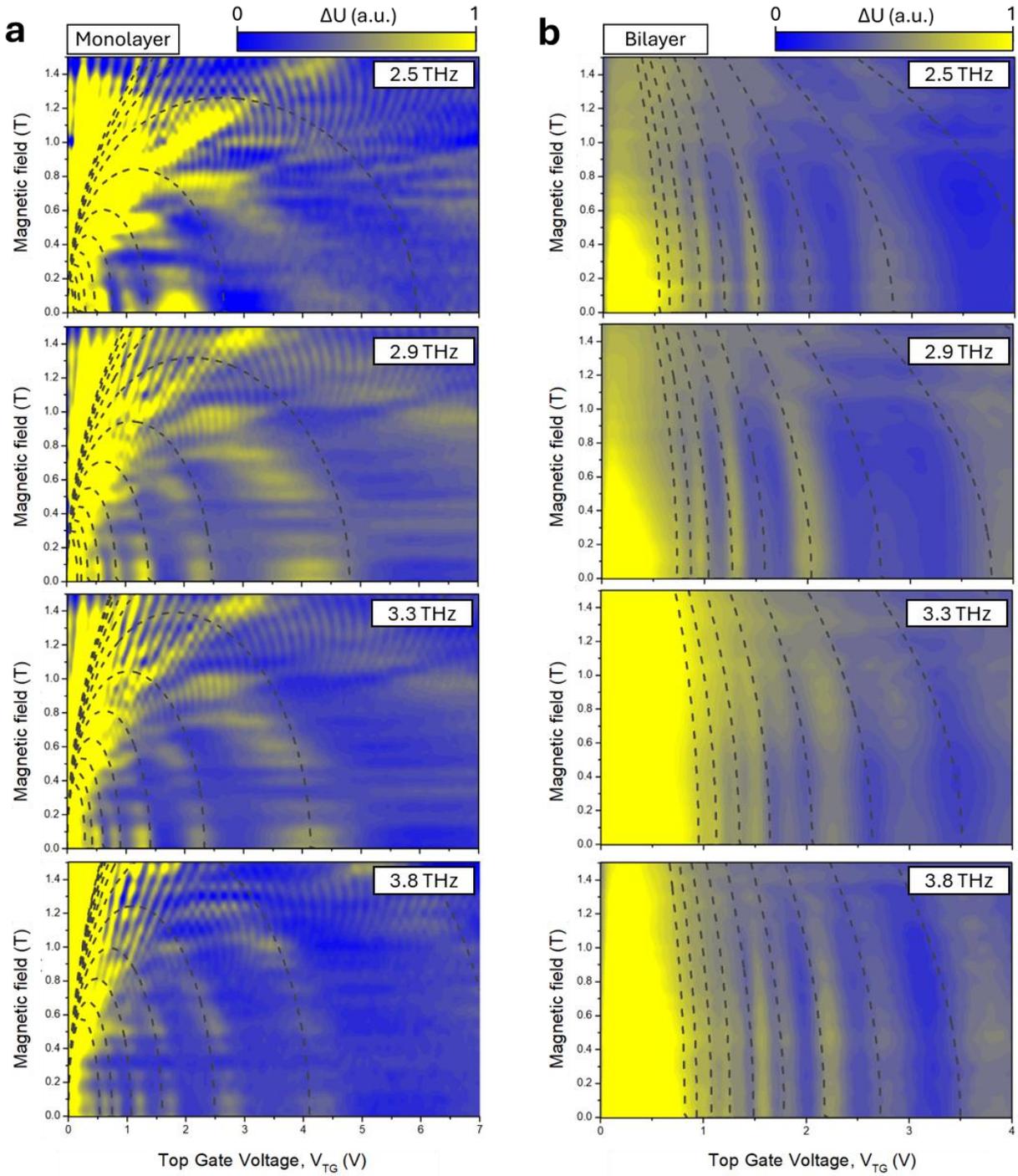

**Figure S8. Frequency dependence of magnetoplasmon.** Mapping of the THz photoresponse as a function of magnetic field and the applied top-gate voltage under excitation frequencies of 2.5, 2.9, 3.3 and 3.8 THz measured at 1.7 K for monolayer graphene, **a**, and at 15K for bilayer graphene, **b**. Theoretical curves for the magnetoplasmon resonance positions are shown in gray dashed lines, calculated using Eq. (10) for the Dirac (monolayer) system and Eq. (11) for the Schrödinger (bilayer) system, as detailed in the main manuscript. A good qualitative agreement is observed between the experimental data and the respective theoretical predictions for both systems.

# REFERENCES

(1) Zomer, P. J.; Guimarães, M. H. D.; Brant, J. C.; Tombros, N.; van Wees, B. J. Fast Pick up Technique for High Quality Heterostructures of Bilayer Graphene and Hexagonal Boron Nitride. *Appl Phys Lett* **2014**, *105* (1), 013101. https://doi.org/10.1063/1.4886096.

(2) Delgado-Notario, J. A.; Power, S. R.; Knap, W.; Pino, M.; Cheng, J.; Vaquero, D.; Taniguchi, T.; Watanabe, K.; Jesús, J.; Velázquez-Pérez, E.; Meziani, Y. M.; Alonso-González, P.; Caridad, J. M. Unveiling the Miniband Structure of Graphene Moiré Superlattices via Gate-Dependent Terahertz Photocurrent Spectroscopy. *ACS Nano* **2025**. https://doi.org/10.1021/ACSNANO.5C05306.

(3) Domaratskiy, I. K.; Kashchenko, M. A.; Semkin, V. A.; Mylnikov, D. A.; Titova, E. I.; Svintsov, D. A. Natural Edge Bilayer Graphene Transistor. *Russian Microelectronics 2023 52:1* **2024**, *52* (1), S2–S5. https://doi.org/10.1134/S1063739723600541.

(4) Chou, S. Y.; Antoniadis, D. A.; Smith, H. I. Application of the Shubnikov-de Haas Oscillations in the Characterization of Si MOSFET's and GaAs Modfet'S. *IEEE Trans Electron Devices* **1987**, *34* (4), 883–889. https://doi.org/10.1109/T-ED.1987.23011.

(5) Caridad, J. M.; Castelló, Ó.; López Baptista, S. M.; Taniguchi, T.; Watanabe, K.; Roskos, H. G.; Delgado-Notario, J. A. Room-Temperature Plasmon-Assisted Resonant THz Detection in Single-Layer Graphene Transistors. *Nano Lett* **2024**, *24* (3), 935–942. https://doi.org/10.1021/acs.nanolett.3c04300.

(6) Bandurin, D. A.; Svintsov, D.; Gayduchenko, I.; Xu, S. G.; Principi, A.; Moskotin, M.; Tretyakov, I.; Yagodkin, D.; Zhukov, S.; Taniguchi, T.; Watanabe, K.; Grigorieva, I. V; Polini, M.; Goltsman, G. N.; Geim, A. K.; Fedorov, G. Resonant Terahertz Detection Using Graphene Plasmons. *Nat Commun* **2018**, *9* (1), 5392. https://doi.org/10.1038/s41467-018-07848-w.

(7) Satou, A.; Ryzhii, V.; Mitin, V.; Vagidov, N. Damping of Plasma Waves in Two-Dimensional Electron Systems Due to Contacts. *physica status solidi (b)* **2009**, *246* (9), 2146–2149. https://doi.org/10.1002/PSSB.200945269.

(8) Shchepetov, A.; Gardès, C.; Roelens, Y.; Cappy, A.; Bollaert, S.; Boubanga-Tombet, S.; Teppe, F.; Coquillat, D.; Nadar, S.; Dyakonova, N.; Videlier, H.; Knap, W.; Seliuta, D.; Vadoklis, R.; Valušis, G. Oblique Modes Effect on Terahertz Plasma Wave Resonant Detection in InGaAs∕InAlAs Multichannel Transistors. *Appl Phys Lett* **2008**, *92* (24), 242105. https://doi.org/10.1063/1.2945286.

(9) Svintsov, D. Hydrodynamic-to-Ballistic Crossover in Dirac Materials. *Phys Rev B* **2018**, *97* (12), 121405. https://doi.org/10.1103/PhysRevB.97.121405.

(10) Boubanga-Tombet, S.; Teppe, F.; Coquillat, D.; Nadar, S.; Dyakonova, N.; Videlier, H.; Knap, W.; Shchepetov, A.; Gardès, C.; Roelens, Y.; Bollaert, S.; Seliuta, D.; Vadoklis, R.; Valušis, G. Current Driven Resonant Plasma Wave Detection of Terahertz Radiation: Toward the Dyakonov–Shur Instability. *Appl Phys Lett* **2008**, *92* (21), 212101. https://doi.org/10.1063/1.2936077.